\documentclass[prl, twocolumn]{revtex4-1} 

\usepackage{hyperref}
\usepackage{amsmath}
\usepackage{balance}
\usepackage{graphicx}
\usepackage{color}
\usepackage{setspace}

\hypersetup{
    bookmarks=true,         
    unicode=false,          
    pdftoolbar=true,        
    pdfmenubar=true,        
    pdffitwindow=false,     
    pdftitle={},    
    pdfauthor={},     
    pdfsubject={},   
    pdfcreator={},   
    pdfproducer={},  
    pdfkeywords={}, 
    pdfnewwindow=true,      
    colorlinks=false,       
    linkcolor=red,          
    citecolor=green,        
    filecolor=magenta,      
    urlcolor=cyan           
}

\pdftrailerid{} 
\begin{document}

\title{The pure-quartic soliton laser}


\author{Antoine F. J. Runge$^{1,*}$}
\author{Darren D. Hudson$^{2}$}
\author{Kevin K. K. Tam$^{1}$}
\author{C. Martijn de Sterke$^{1,3}$}
\author{Andrea Blanco-Redondo$^{4}$}

\affiliation{
$^{1}$Institute of Photonics and Optical Science (IPOS), School of Physics, The University of Sydney, NSW 2006, Australia\\
$^{2}$MQ Photonics, Department of Physics and Astronomy, Macquarie University, NSW 2109, Australia\\
$^{3}$The University of Sydney Nano Institute (Sydney Nano), The University of Sydney, NSW 2006, Australia\\
$^{4}$Nokia Bell Labs, 791 Holmdel Road, Holmdel, NJ 07733, USA\\
$^{*}$Corresponding author: antoine.runge@sydney.edu.au
}

\begin{abstract}
The generation of ultrashort pulses hinges on the careful management of dispersion. Traditionally, this has exclusively involved second-order dispersion, while higher-order dispersion was treated as a nuisance to be minimized. Here we show that high-order dispersion can be strategically leveraged to access an uncharted regime of ultrafast laser operation. In particular, we demonstrate a mode-locked laser, with an intra-cavity spectral pulse-shaper, that emits pure-quartic soliton pulses, which arise from the interaction of the fourth-order dispersion and the Kerr nonlinearity. Using phase-resolved measurements we demonstrate that the energy of these pulses is proportional to the third power of the inverse pulse duration. This implies a dramatic increase in the energy of ultrashort pulses compared to those emitted by soliton lasers to date. These results not only demonstrate a novel approach to ultrafast lasers, but more fundamentally, they clarify the use of higher-order dispersion for optical pulse control, opening up a plethora of possibilities in nonlinear optics and its applications.
\end{abstract}

\maketitle

\section*{Introduction}

Ultrafast lasers have been fundamental for the development of major photonic applications such as telecommunications \cite{Mollenauer_1991, Mollenauer2_1991, Haus_1996}, supercontinuum \cite{Husakou_2001, Dudley_2006}, nonlinear imaging \cite{Xu_2013} and frequency comb generation \cite{Cundiff_2003}. Soliton effects, based on the balance of quadratic dispersion and nonlinearity, have allowed for the direct generation of optical pulses with duration below 10~fs \cite{Zhou_1994, Jung_1997}. Soliton fibre lasers, which have relatively simple configurations can be built using off-the-shelf components \cite{Mollenauer_1984, Kafka_1989, Matsas_1992}. However, the energy of the pulses emitted by these lasers is limited by the soliton area theorem \cite{Zakharov_1972, Hasegawa_1973}, and by the appearance of resonant spectral sidebands, arising from periodic perturbations in the optical cavity \cite{Kelly_1992}.

Pure-Quartic Solitons (PQSs), shape-maintaining pulses that arise from the balance of the Kerr nonlinearity and negative fourth-order dispersion (FOD), were recently discovered in dispersion-engineered photonic crystal waveguides \cite{Blanco_Redondo_2016}. They were observed in a spectral range where the second-order (quadratic) dispersion $\beta_2$ was positive, the third-order (cubic) dispersion $\beta_3$ was negligible and the fourth-order (quartic) dispersion $\beta_4$ was negative. Recent theoretical studies have unveiled that PQSs possess an advantageous energy scaling \cite{Lo_2018, Tam_2019}, which grants them the potential to achieve significantly higher energy than their conventional soliton counterpart for short pulse durations. This discovery has led to efforts to transition from planar silicon photonic crystals to other platforms \cite{Lo_2018, TaheriMatsko_2019}, including optical fibres where mature fibre laser technology can be utilized. Initial approaches involved carefully designed photonic crystal fibre (PCF) with multi-layer air holes that exhibit negligible $\beta_2$ and large, negative $\beta_4$ \cite{Lo_2018}. However, the fabrication tolerances required for achieving the necessary control of dispersion are very tight. Consequently, PQSs have until now not been observed experimentally in optical fibres.

Here, we report the first experimental demonstration of a mode-locked fibre laser emitting pure-quartic soliton pulses. We achieve the necessary laser cavity dispersion conditions by utilizing an intra-cavity spectral pulse shaper that simultaneously cancels the $\beta_2$ and $\beta_3$ of the otherwise conventional-fibre cavity, while imparting a strong, negative $\beta_4$. PQSs arise from this lumped quartic dispersion in much the same way that conventional solitons can arise in laser cavities with discrete segments of normal and anomalous dispersion fibre \cite{Knox_1994,Turitsyn_2012}.

Using spectral, temporal and phase-resolved measurements we demonstrate that the output pulses behave significantly differently than any other type of laser pulses: we find that the energy of the emitted pulses $E_{\rm PQS}$ is proportional to $\tau^{-3}$, consistent  with theoretical predictions \cite{Tam_2019}, and in contrast to conventional solitons for which  $E\propto \tau^{-1}$ \cite{Agrawal_NFO}. We also find that the output spectra exhibit sidebands, which we associate with resonant dispersive waves. Though these sidebands also occur in conventional soliton lasers \cite{Kelly_1992},  the measured sideband frequencies are characteristic for cavities with quartic dispersion, and obey a novel equation that we present here. These results constitute strong evidence for a novel operating regime for mode-locked lasers, opening the way for the generation of high-energy, ultrashort optical pulses, arising from SPM and higher-order cavity dispersion.

\section*{Results}

\begin{figure}[ht]
\centering
\includegraphics[width=88mm,clip = true]{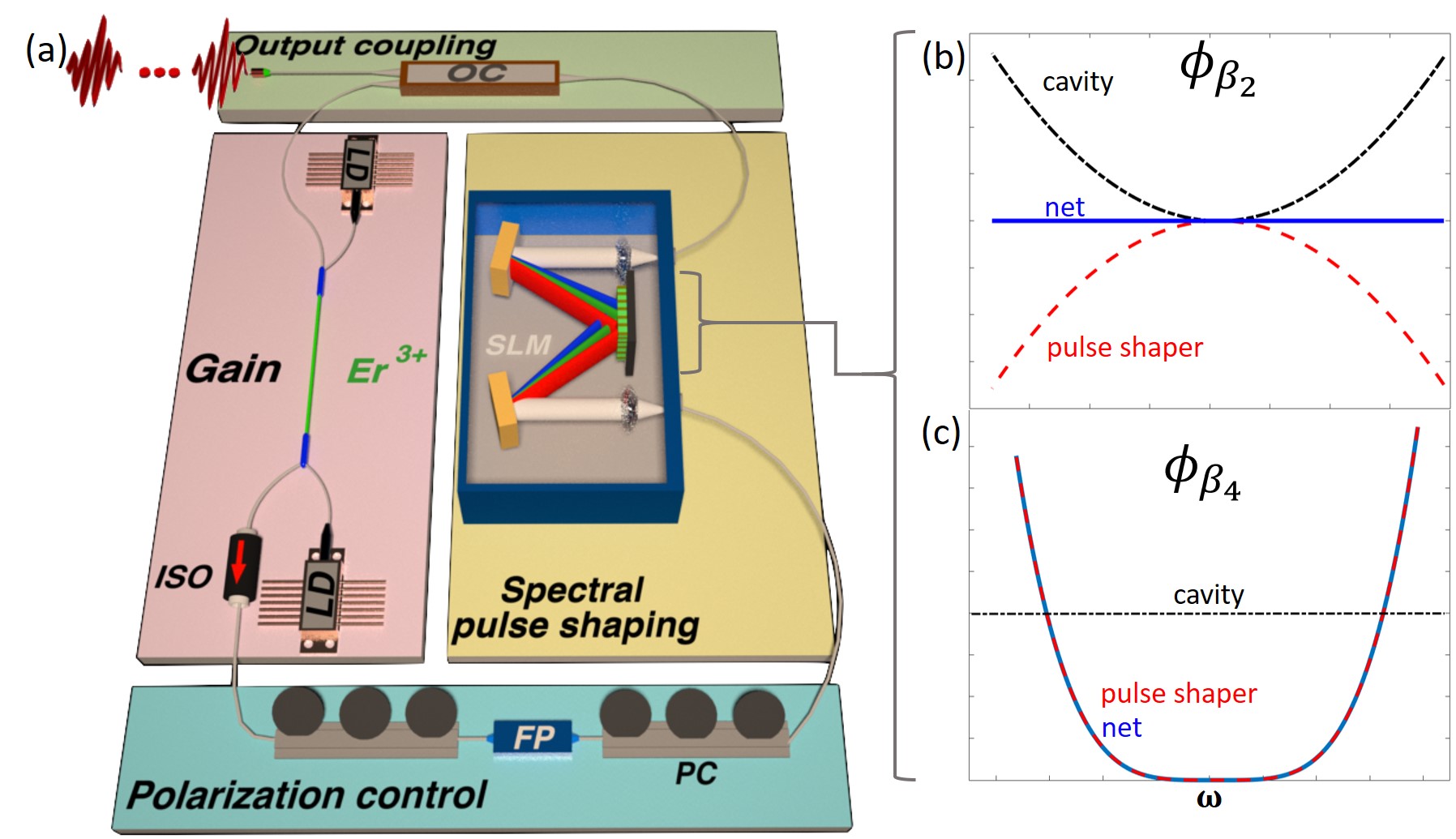}
\vskip-0mm
\caption{Principle of operation of the pure-quartic soliton laser. (a) Schematic of the erbium-doped laser cavity with the following components: Er$^{3+}$, Erbium-doped fibre; LD, 976 nm laser diode; FP, in-line fibre polarizer; PC, polarisation controller; SLM, spatial light modulator; OC, output coupler. (b) Quadratic and (c) quartic dispersive phase imparted by the cavity (dash-dot black) and the spectral pulse-shaper (dash red), with the net quadratic and quartic phase shown in blue. }
\label{laser}
\vskip-1mm
\end{figure}

Our laser configuration is shown schematically in Fig.~\ref{laser}(a). It is an erbium-doped fibre laser using nonlinear polarization evolution (NPE) for mode-locking, and incorporating an intracavity programmable optical pulse-shaper \cite{Schroder_2010, Peng_2016}. The total cavity length is approximately $L=21.4~{\rm m}$, leading to a fundamental repetition rate of 9.3~MHz. Mode-locking is achieved by adjusting the pump power and the polarization controllers (PCs). An output coupler (OC) is located after the fibre polarizer and extracts 50\% of the intracavity power. The pulse-shaper, based on a spatial light modulator (SLM), can produce an arbitrary phase mask and is used to adjust the net cavity dispersion as illustrated in Figs.~\ref{laser}(b) and (c).

\begin{figure*}[hbt]
\centering
\includegraphics[width=\textwidth,clip = true]{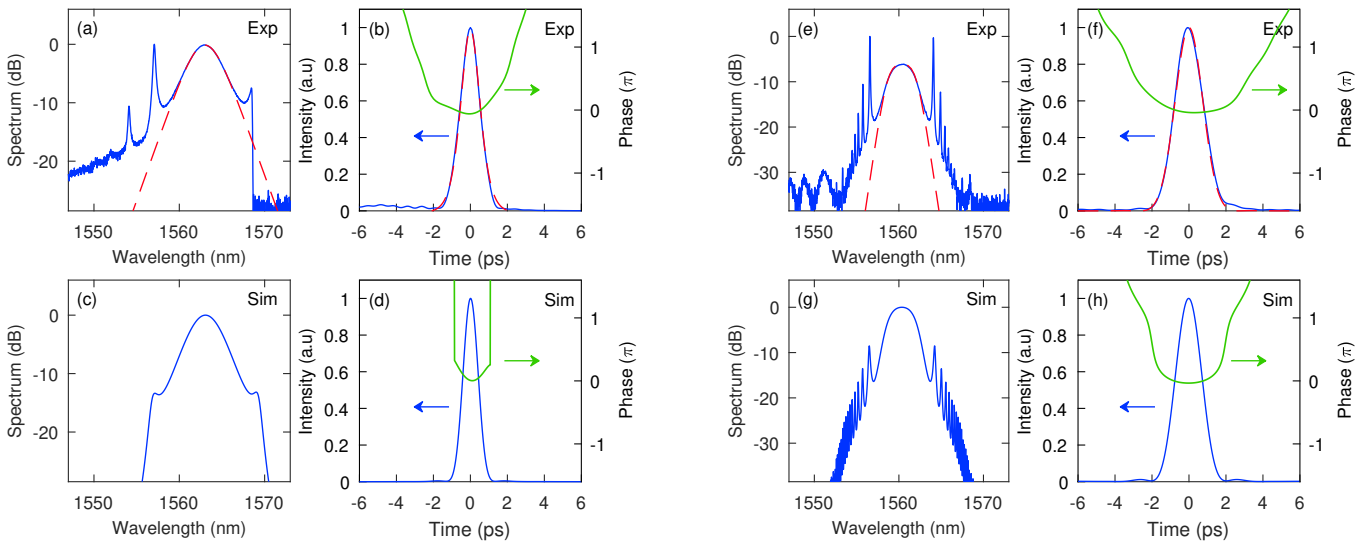}
\vskip-3mm
\caption{Measured and simulated output characteristics of the laser in the conventional soliton (a), (b), (c), (d) and PQS (e), (f), (g), (h) regimes. (a), (e) Measured output spectrum (solid blue curve). Red dashed curves show the corresponding soliton and PQS spectral shapes. (b), (f), retrieved (solid blue curve) and theoretically predicted (red dashed curve) temporal intensity profiles. The solid green curve are the recovered temporal phase. (c), (g), simulated output spectra. (d), (h), simulated intensity profiles (blue) and phases (green).}
\label{spectrum}
\vskip-1mm
\end{figure*}

When no phase mask is applied, the laser operates in the conventional soliton regime, with the intrinsic anomalous GVD of the fibre balancing the SPM. The output pulses display the well-known hyperbolic secant spectrum centred around 1563~nm with a 3.72~nm bandwidth at $-3~{\rm dB}$, and strong spectral (``Kelly'') sidebands \cite{Kelly_1992} as seen in Fig.~\ref{spectrum}(a) (solid blue curve). The truncated long wavelength edge is due to the finite spectral bandwidth of the pulse shaper. To gain insight in the temporal properties of the output pulses we also perform a set of temporal and phase-resolved measurements using a frequency-resolved electrical gating (FREG) setup \cite{Blanco_Redondo_2016, Dorrer_2002}. The retrieved pulse of the laser operating in conventional soliton regime is shown in Fig.~\ref{spectrum}(b) (blue solid line).

We then program the pulse-shaper to induce a phase profile that can be written as
\begin{equation}
   \phi = \exp\left(iL\sum{\beta_n\omega^n/n!}\right).
\label{phase_eq}
\end{equation}
where $\beta_n$ is the $n^{th}$ dispersion order for $n=2,3,4$. For the results discussed below $\beta_2 = +21.4~{\rm ps^2/km}$, $\beta_3 = -0.12~{\rm  ps^3/km}$, and we take initially $\beta_4 = -80~{\rm ps^4/km}$. The pulse shaper serves two aims; (i) to compensate the $2^{\rm nd}$ and $3^{\rm rd}$ order dispersion introduced by the optical fibres \cite{Hammani_2011}, so that net cavity dispersion of these orders is negligible (Fig.~\ref{laser}(b)); and (ii) the pulse shaper generates a large negative FOD, which dominates the propagation (Fig.~\ref{laser}(c)). The applied FOD is several orders of magnitude larger than the intrinsic FOD of single mode fibres ($\approx10^{-3}$ ps$^4$/km) \cite{Ito_2016}. The output spectrum of the laser operating in this regime, shown in Fig.~\ref{spectrum}(e), has a 3.16~nm bandwidth with a shape that differs significantly from that of conventional solitons, but is in very good agreement with the theoretically predicted spectral profile of the PQS (red dashed line) \cite{Tam_2019}. We note in particular the PQS's distinct flatness of the spectral maximum. We also observe several strong, narrowly spaced spectral sidebands. The retrieved pulse corresponding to this spectrum is shown in Fig.~\ref{spectrum}(f) (blue solid line). The recovered pulse has a duration of $\tau = 1.74$~ps (at full half width maximum). The retrieved temporal pulse shape is again in good agreement with the theoretically predicted PQS shape for the same pulse duration (red dashed line) \cite{Tam_2019}. From this measurement and the spectral bandwidth at $-3~{\rm dB}$ measured from the corresponding spectrum, we calculate a time-bandwidth product of $0.67$. This value is larger than the predicted value of $0.53$ for transform-limited PQSs \cite{Tam_2019}, indicating that the pulses at the laser output have a nearly flat phase with a modest higher-order phase distortion, as suggested by the recovered phase (green solid line in Fig.~\ref{spectrum}(f)). Finally, we note that the predicted temporal shape of the PQS exhibits periodic oscillations in the tails that are not observed in the retrieved experimental profile \cite{Tam_2019}. This is because the first lobe is expected to appear approximately $28$~dB below the pulse's central maximum, which is below the continuous background arising from the dispersive spectral sidebands in our experiments.

We model the laser dynamics using an iterative cavity map \cite{Oktem_2010, Woodward_2018}, in which the propagation through every element is modeled using a generalized nonlinear Schr\"{o}dinger equation with Kerr nonlinearity, and with  dispersion up to the fourth order (see Methods). The simulated spectra and intensity profiles corresponding to the two operating regimes are shown in Fig.~\ref{spectrum}(c), (d), (g) and (h), are in excellent agreement with the experimental results, and are consistent with the change in the pulse-shaping mechanism.

We can assess the quartic nature of the laser by analyzing the spectral positions of the dispersive waves in Fig.~\ref{spectrum}(e). Similar to conventional solitons, these dispersive waves arise from perturbations when the soliton propagates through the cavity; the spectral positions of these peaks provide information about the cavity dispersion \cite{Dennis_1994}. The dispersive waves are generated every round trip and interfere constructively when $\beta_{PQS}-\beta_{\rm lin} =2m\pi/L$, where $\beta_{PQS,{\rm lin}}$ are the propagation constants of the soliton and the dispersive wave, respectively, and $m$ is an integer. When this condition is satisfied, then dispersive waves generated in consecutive passages through the cavity, interfere constructively, leading to narrow spectral peaks \cite{Kelly_1992}.

\begin{figure*}[hbt]
\centering
\includegraphics[width=\textwidth,clip = true]{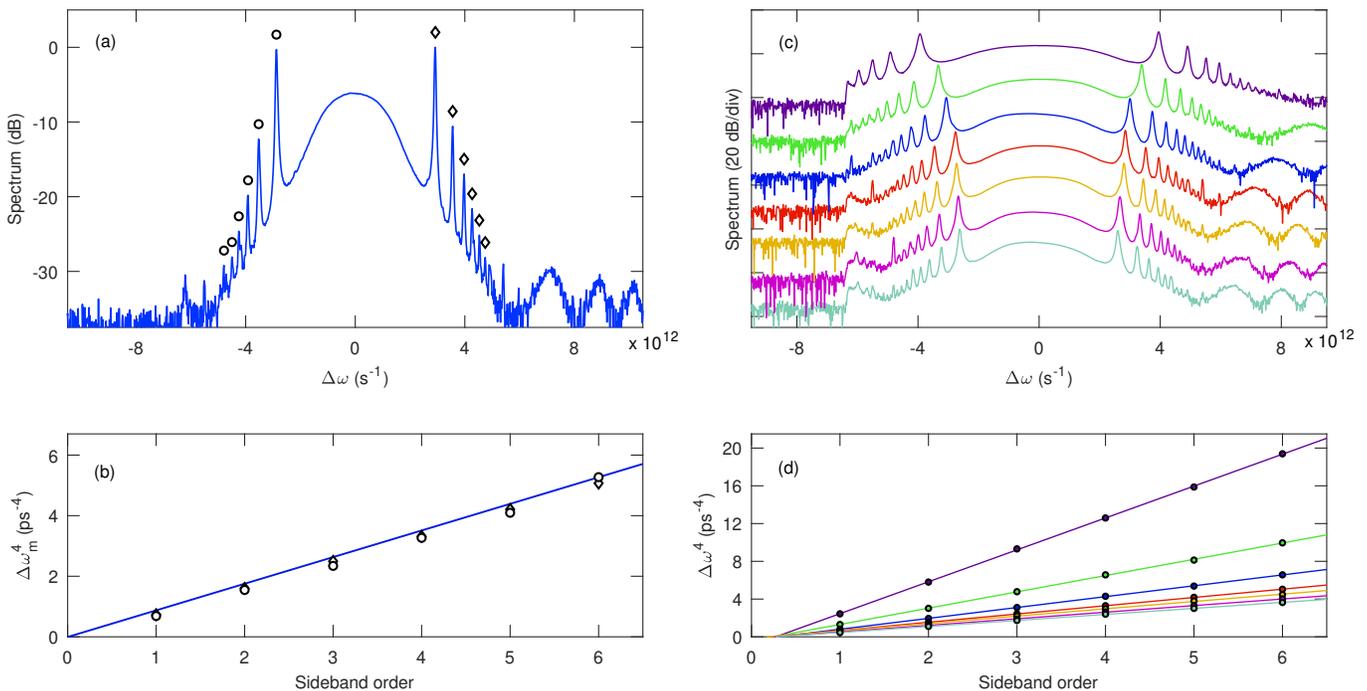}
\caption{(a) Measured spectrum centered around $\omega_0$. The circles mark the measured spectral positions of the sidebands for the low (circles) and high (diamonds) frequencies. (b) Fourth power of the of the sidebands positions as a function of the sideband order. The solid blue line shows the nominal spectral positions of the sidebands calculated from Eq.~\ref{sideband_eq}. (c) Measured optical spectra corresponding to the shortest pulse duration recorded for each value of $\beta_4$. From top to botttom, $\beta_4 = -20~{\rm ps^4/km}$ (purple), $-40~{\rm ps^4/km}$ (green), $-60~{\rm ps^4/km}$ (blue), $-80~{\rm ps^4/km}$ (red), $-90~{\rm ps^4/km}$ (yellow), $-100~{\rm ps^4/km}$ (pink) and $-110~{\rm ps^4/km}$ (light blue). (d) Corresponding fourth power of the of the sidebands positions as a function of the sideband order. The solid colour lines correspond to linear fits.}
\label{sidebands}
\end{figure*}

For a linear wave propagating in a quartic dispersion cavity, $\beta_{\rm lin} = -|\beta_4|(\omega-\omega_0)^4/24$, whereas the PQS experiences a constant dispersion across its entire bandwidth of $\beta_{PQS} = K|\beta_4|/\tau^4$, where the
constant $K=1.67$ \cite{Tam_2019}. Using the argument outlined in the previous paragraph, it is straightforward to show that the $m^{th}$-order spectral resonances $\omega_m$ satisfy
\begin{equation}
   \omega_m = \pm{1\over\tau} \left({48m\pi\tau^4\over |\beta_4|L}-24K\right)^{1/4}.
\label{sideband_eq}
\end{equation}
This equation shows that the $4^{th}$ power of the sideband frequency offsets are equally spaced by $48\pi/(|\beta_4|L)$. In Fig.~\ref{sidebands}(a) we show the optical spectrum of the PQS centred around $\omega_0$ and we mark the measured spectral positions of the sidebands for the low (circles) and high (diamonds) frequencies of the spectrum. Figure \ref{sidebands}(b) shows the $4^{th}$ power of these sidebands positions ($\Delta\omega^4$) versus the sideband order. This confirms that they follow a linear relation, consistent with Eq.~\ref{sideband_eq}. The spacing determined from the measured spectral position of the sidebands is fixed at $0.90~{\rm ps^{-4}}$ and $0.867~{\rm ps^{-4}}$ for the low (circles) and high (diamonds) frequencies, respectively. These results are in excellent agreement with the theoretical value of $0.88~{\rm ps^{-4}}$ following from Eq.~\ref{sideband_eq}. The small discrepancies between the calculated and theoretical values could be due to the limited resolution of the measurement of the position of the sidebands, or to residual, uncompensated quadratic and cubic dispersion.

\begin{figure*}[ht]
\centering
\includegraphics[height=10.5cm,clip = true]{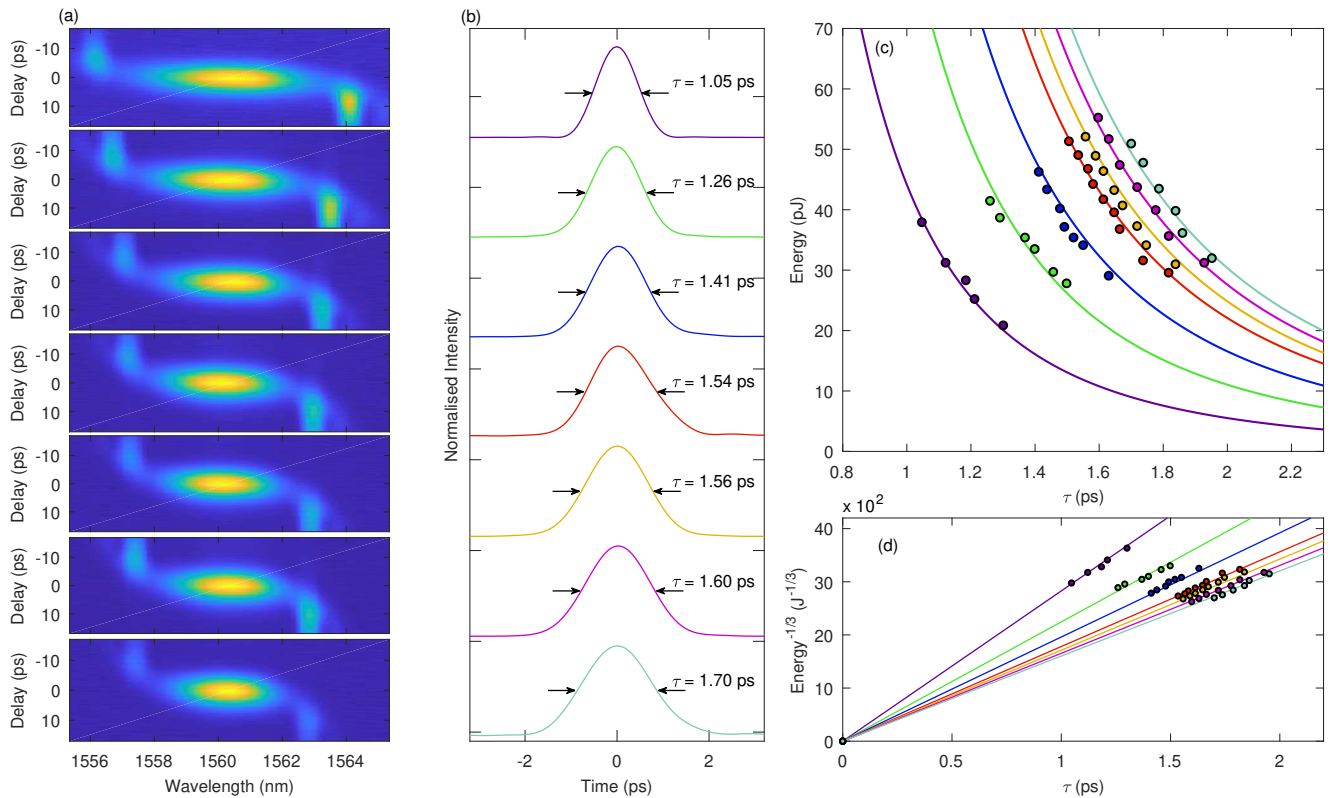}
\vskip-3mm
\caption{Measured energy-width scaling properties of the emitted PQS pulses for different values of applied quartic dispersion. (a) Measured spectrograms for the shortest pulse durations recorded for each value of $\beta_4$. From top to botttom, $\beta_4 = -20~{\rm ps^4/km}$ (purple), $-40~{\rm ps^4/km}$ (green), $-60~{\rm ps^4/km}$ (blue), $-80~{\rm ps^4/km}$ (red), $-90~{\rm ps^4/km}$ (yellow), $-100~{\rm ps^4/km}$ (pink) and $-110~{\rm ps^4/km}$ (light blue). (b) Corresponding retrieved temporal intensity profiles. (c) Measured pulse energy $E$ versus pulse duration (circles). The solid curves are $\propto\beta_4\tau^{-3}$, as follows from  Eq.~\ref{energy}. (d) $E^{-1/3}$ versus pulse duration shown over a wider range.}
\label{E_vs_tau}
\vskip-1mm
\end{figure*}

While the results discussed thus far were taken at a fixed $\beta_4=-80~{\rm ps^4/km}$, in Figs~\ref{sidebands}(c) and (d) we show, for completeness, similar measurements for seven different values of the quartic dispersion, namely $\beta_4=-20, -40, -60, -80,$ $
-90, -100, -110~{\rm ps^4/km}$. Fig.~\ref{sidebands}(c) shows measured spectra for the shortest pulses that we generated at each of these $\beta_4$ values (see Fig.~\ref{E_vs_tau}(b)). The similarity of these spectra implies that the pulse shaping mechanism is unchanged for all the $\beta_4$ values considered. Similarly, in Fig.~\ref{sidebands}(d) we show an analysis of the positions of the side bands. Since for all cases the $\Delta\omega^4$ values lie on straight lines, we confirm that the dispersion remains quartic for all $\beta_4$ values. The presence of the last term in the brackets in Eq.~\ref{sideband_eq} implies that the straight lines do not go through the origin, or even go through a common point.

We now investigate in more detail the results of FREG measurements as we vary $\beta_4$.
In Fig.~\ref{E_vs_tau}(a) we show spectrograms of the shortest pulses measured, for each $\beta_4$. The corresponding retrieved temporal intensity profiles are shown in Fig.~\ref{E_vs_tau}(b). Similar to conventional soliton lasers, reducing the net-cavity dispersion allows for the generation of shorter optical pulses \cite{Tamura_1993, Chen_1999, Turitsyn_2012}. The vertical streaks at short and long wavelengths in Fig.~\ref{E_vs_tau}(a) are the first sidebands on either side of the pulse spectrum, discussed earlier. As required for negative quartic dispersion, on the short wavelength side they precede the pulse, whereas on the long wavelength side they follow it.

Finally, we report the energy-width scaling of the emitted PQSs. While for conventional solitons the pulse energy is inversely proportional to the pulse duration, recent theoretical and numerical studies show that for a PQS the pulse energy is given by \cite{Tam_2019}
\begin{equation}
    E_{PQS} = \frac{2.87|\beta_4|}{\gamma\tau^3}.
    \label{energy}
\end{equation}
where $\gamma$ is the average cavity nonlinear parameter. At each value of the quartic dispersion we measure the spectral bandwidth of the output pulses for different pulse energies by adjusting the pump power, and we deduct the portion of the energy in the spectral sidebands by integrating the measured optical spectrum. The corresponding pulse durations are then calculated using the time-bandwidth product of $0.67$ determined above, which we find to be constant for the entire range of parameters that we consider.
The results of the procedure described in the previous paragraph are summarized in  Figs~\ref{E_vs_tau}(c) and (d). The circles in Figure~\ref{E_vs_tau}(c) show the measured pulse energies versus the pulse duration $\tau$, for different values of quartic dispersion. The results in Fig.~\ref{E_vs_tau}(c) are in excellent agreement with Eq.~\ref{energy} once we account for the output coupling and the variations of the pulse parameters within the cavity (see Methods). To see this more clearly, we plot $E\propto\beta_4\tau^{-3}$ for each of the $\beta_4$ values. This shows that for fixed $\beta_4$, the energy $E\propto \tau^{-3}$ and that for fixed $\tau$, the energy $E\propto\beta_4$, consistent with Eq.~\ref{energy}.

Figure~\ref{E_vs_tau}(d) shows the same data as Fig.~\ref{E_vs_tau}(c), but has $E^{-1/3}$ on the vertical axis. Plotted in this way, the data should form a fan of straight lines, with each rib, corresponding to a particular value of $\beta_4$, going through the origin. The consistency between the measured data and the prediction from Eq.~\ref{energy} is conclusive evidence of the unique scaling properties of the pulses emitted by our laser.

A second set of measurements (not shown here) is performed with a different output coupler, extracting 10\% of the intracavity power, and we obtain similar results. We also perform a similar measurement (see Supplementary information) for the laser operating in the conventional soliton regime (as seen in Fig.~\ref{spectrum}(a)) and find the output pulse energy follows the well-known relation $E\propto 1/\tau$ \cite{Agrawal_NFO}. These results confirm that PQS pulses follow a different energy-width scaling relation and that they could outperform conventional soliton for short pulse durations, as suggested by previous studies \cite{Tam_2019, Blanco_Redondo_2017}.

\section*{Discussion}
By harnessing the interaction between fourth-order dispersion with the nonlinearity in a fibre laser cavity we provide the first experimental realization of a pure-quartic soliton laser. We have provided experimental evidence of a novel pulse spectrum and temporal shape, energy-width scaling, and spectral sideband behaviour, with respect to any type of laser pulses produced to date.

The necessary FOD-dominating dispersion profile was achieved by incorporating a reconfigurable spectral pulse shaper inside the laser cavity. This innovation is crucial to leverage the effects of high-order dispersion in ultrafast lasers, without the need for challenging dispersion engineering of the waveguide or fibre.

The range of pulse energies achievable with our current setup is limited by the spectral and delay range of the spectral pulse-shaper. This limitation is in no way fundamental and it could be overcome by using a more broadband pulse-shaper with a larger dynamic range. In the future, fibre platforms with intrinsically high negative FOD \cite{Lo_2018} could substitute the pulse shaper. Alternatively, microresonator geometries with dominant negative FOD have been recently theoretically studied \cite{TaheriMatsko_2019}, which highlights a potential route towards laser systems leveraging high-order dispersion in integrated platforms.

Our results could pave the way for a new class of simple and low-cost mode-locked fibre laser emitting ultrashort high-energy pulses. The simple laser cavity configuration presented in this work offers an exceedingly flexible testbed for the generation and the study of optical pulses arising from the interplay of Kerr nonlinearity and hybrid dispersion – combination of different orders of dispersion, including orders higher than fourth. This provides new degrees of freedom to optically control the shape and the energy scaling of optical pulses which could not only have impact on only ultrafast lasers, but also on other areas in which these features are crucial such as frequency combs, supercontinuum generation, and advanced modulation formats for communications.

\section*{Methods}
\textbf{Numerical simulation model.} Numerical simulations are based on the nonlinear Schr\"{o}dinger equation:
\begin{equation}
   \frac{\partial A}{\partial z}+i\hat{D}\left(i\frac{\partial}{\partial T}\right)A=\frac{g}{2}A+i\gamma A|A|^2.
    \label{GNLSE}
\end{equation}
Here, $A = A(z,T)$ is the slowly varying amplitude of the pulse envelope, $z$ is the propagation coordinate, $T$ is the pulse local time, and $\gamma$ is the nonlinear parameter given by $\gamma = n_2\omega_0/(cA_{eff})$, with $n_2$ is the nonlinear refractive index, $\omega_0$ is the central frequency , $c$ is the speed of light in vacuum, and $A_{eff}$ is the effective mode area. The dispersion operator is defined as:
\begin{equation}
\hat{D}\left(i\frac{\partial}{\partial T}\right) = \sum_{k}\frac{\beta_k}{k!}\left(i\frac{\partial}{\partial T}\right)^k,
    \label{Dop}
\end{equation}
for $k = 2,3,4$. $\beta_k$ is the $k^{th}$ order of dispersion. The gain in the doped fibre section is calculated using:
\begin{equation}
   g = \frac{g_0}{1+E(z)/E_{sat}}.
    \label{gain}
\end{equation}
where $g_0 = 3.45$ is the small-signal gain (corresponding to 30 dB in power and non-zero only in the doped-fibre section), $E(z) = \int|A(z,T)|^2$ is the pulse energy and $E_{sat}$ is the saturation energy, which is adjusted to simulate changing the pump power. We multiply $g(z)$ with a Lorentzian profile of $50$ nm width to form the finite gain bandwidth $g(z,\omega_0)$. The saturable absorber is modelled by a transfer function that describes its transmittance
\begin{equation}
   T(\tau) = 1-\frac{q_0}{1+P(\tau)/P_0}.
    \label{SA}
\end{equation}
where $q_0$ is the unsaturated loss of the saturable absorber, $P(\tau) = |A(z,\tau)|^2$ is the instantaneous pulse power and $P_0$ is the saturation power. The spectral pulse-shaper is modelled by multiplying the electric field by a phase following the expression in Eq.~\ref{phase_eq} in the spectral domain. The insertion losses ($\approx 5.6$ dB) of the spectral pulse-shaper are also taken into account in the simulations. Our numerical model is solved with a standard symmetric split-step Fourier method algorithm. The dispersion operator is applied in the frequency domain, whilst the nonlinear terms and gain are calculated in the time domain. Four our simulations we have used an initial field composed of Gaussian random noise multiplied by a sech shape in the time domain. The same stable solutions are reached for different initial noise fields.

The parameters used in our numerical simulations are the same as their experimental values. The erbium-doped fibre is $1.5$ m long, with a mode-field diameter (MFD) of $9.5$ $\mu$m, numerical aperture (NA) of 0.13 and $\gamma = 0.0016$ W$^{-1}$m$^{-1}$ at $1560$ nm. The rest of the cavity is built from SMF-28 fibre. SMF-28 has an MFD of $10.4$ $\mu$m, NA of 0.14, $\gamma = 0.0013$ W$^{-1}$m$^{-1}$ at $1560$ nm. The dispersion coefficient are, $\beta_2 = -21.4$ ps$^{2}$km$^{-1}$, $\beta_3 = 0.12$ ps$^{3}$km$^{-1}$ and $\beta_4 = -0.0022$ ps$^{4}$km$^{-1}$. For the simulated results shown in Fig.~\ref{spectrum}, $q_0 = 0.7$ and $P_0 = 500 W$. The saturation energy was set at $E_{sat} = 90$ pJ and $E_{sat} = 100$ pJ, for the conventional and pure-quartic solitons, respectively.

\textbf{Mode-locking.} $980$ nm light from two laser diodes is delivered to the cavity through two $980/1550$ nm wavelength division multiplexers. An optical isolator is used to ensure unidirectional pulse propagation in the cavity. Nonlinear polarisation evolution is implemented using a set of two fibre polarisation controllers and a fibre polariser, which act as an artificial saturable absorber. The laser is self-starting and multi-pulsing at a pump power of $370$ mW after adjusting the polarisation controller. Single pulsing is then achieved by decreasing the pump power to $155$ mW.

\textbf{Phase-resolved characterization method.} The output pulses were input into the FREG apparatus. The pulses were split into to two branches by a 70/30 fibre-coupler. 30\% of the output power was sent to a branch with a variable delay, before being detected by a fast photodiode and transferred to the electrical domain. This electrical signal drove a drove a Mach–Zender modulator that gated the optical pulses from the 70\% branch of the fibre-coupler. Using an optical spectrum analyser (OSA), we measured the spectra as a function of the delay to generate a series of optical spectrograms. We then de-convolved the spectrograms with a blind deconvolution numerical algorithm (512x512 grid-retrieval errors < 0.005) to retrieve the pulse intensity and the phase in the temporal domain. We verified the validity of the retrieved pulses by taking their Fourier transform and checked these matched with the output spectra measured with the OSA.

\section*{Funding Information}
Australian Research Project (ARC) Discovery Project (DP180102234); University of Sydney \textit{Professor Harry Messel Research Fellowship}; Asian Office of Aerospace R\&D (AOARD) grant (FA2386-19-1-4067).

\section*{Author contributions}

A. B.-R , C. M. d. S, and D. D. H. conceived the idea of the pure-quartic soliton laser. A. F. J. R , D. D. H, and A. B. -R. designed the experiment. A. F. J. R. performed the experiments and the numerical simulations. K. K. K. T and C. M. d. S. carried out the theoretical analysis. C. M. d. S. and A. B.-R supervised the overall project. All the authors contributed to the interpretation of data and wrote the manuscript.

\section*{References}

\bibliography{PQS_bib}

\providecommand{\noopsort}[1]{}\providecommand{\singleletter}[1]{#1}%
\begin{thebibliography}{33}%
\makeatletter
\providecommand \@ifxundefined [1]{%
 \@ifx{#1\undefined}
}%
\providecommand \@ifnum [1]{%
 \ifnum #1\expandafter \@firstoftwo
 \else \expandafter \@secondoftwo
 \fi
}%
\providecommand \@ifx [1]{%
 \ifx #1\expandafter \@firstoftwo
 \else \expandafter \@secondoftwo
 \fi
}%
\providecommand \natexlab [1]{#1}%
\providecommand \enquote  [1]{``#1''}%
\providecommand \bibnamefont  [1]{#1}%
\providecommand \bibfnamefont [1]{#1}%
\providecommand \citenamefont [1]{#1}%
\providecommand \href@noop [0]{\@secondoftwo}%
\providecommand \href [0]{\begingroup \@sanitize@url \@href}%
\providecommand \@href[1]{\@@startlink{#1}\@@href}%
\providecommand \@@href[1]{\endgroup#1\@@endlink}%
\providecommand \@sanitize@url [0]{\catcode `\\12\catcode `\$12\catcode
  `\&12\catcode `\#12\catcode `\^12\catcode `\_12\catcode `\%12\relax}%
\providecommand \@@startlink[1]{}%
\providecommand \@@endlink[0]{}%
\providecommand \url  [0]{\begingroup\@sanitize@url \@url }%
\providecommand \@url [1]{\endgroup\@href {#1}{\urlprefix }}%
\providecommand \urlprefix  [0]{URL }%
\providecommand \Eprint [0]{\href }%
\providecommand \doibase [0]{http://dx.doi.org/}%
\providecommand \selectlanguage [0]{\@gobble}%
\providecommand \bibinfo  [0]{\@secondoftwo}%
\providecommand \bibfield  [0]{\@secondoftwo}%
\providecommand \translation [1]{[#1]}%
\providecommand \BibitemOpen [0]{}%
\providecommand \bibitemStop [0]{}%
\providecommand \bibitemNoStop [0]{.\EOS\space}%
\providecommand \EOS [0]{\spacefactor3000\relax}%
\providecommand \BibitemShut  [1]{\csname bibitem#1\endcsname}%
\let\auto@bib@innerbib\@empty
\bibitem [{\citenamefont {Mollenauer}\ \emph
  {et~al.}(1991{\natexlab{a}})\citenamefont {Mollenauer}, \citenamefont
  {Evangelides},\ and\ \citenamefont {Haus}}]{Mollenauer_1991}%
  \BibitemOpen
  \bibfield  {author} {\bibinfo {author} {\bibfnamefont {L.~F.}\ \bibnamefont
  {Mollenauer}}, \bibinfo {author} {\bibfnamefont {S.~G.}\ \bibnamefont
  {Evangelides}}, \ and\ \bibinfo {author} {\bibfnamefont {H.~A.}\ \bibnamefont
  {Haus}},\ }\href@noop {} {\bibfield  {journal} {\bibinfo  {journal} {J.
  Lightwave Technol.}\ }\textbf {\bibinfo {volume} {9}},\ \bibinfo {pages}
  {194} (\bibinfo {year} {1991}{\natexlab{a}})}\BibitemShut {NoStop}%
\bibitem [{\citenamefont {Mollenauer}\ \emph
  {et~al.}(1991{\natexlab{b}})\citenamefont {Mollenauer}, \citenamefont
  {Neubelt}, \citenamefont {Haner}, \citenamefont {Lichtman}, \citenamefont
  {Evangelides},\ and\ \citenamefont {Nyman}}]{Mollenauer2_1991}%
  \BibitemOpen
  \bibfield  {author} {\bibinfo {author} {\bibfnamefont {L.~F.}\ \bibnamefont
  {Mollenauer}}, \bibinfo {author} {\bibfnamefont {M.~J.}\ \bibnamefont
  {Neubelt}}, \bibinfo {author} {\bibfnamefont {M.}~\bibnamefont {Haner}},
  \bibinfo {author} {\bibfnamefont {E.}~\bibnamefont {Lichtman}}, \bibinfo
  {author} {\bibfnamefont {S.~G.}\ \bibnamefont {Evangelides}}, \ and\ \bibinfo
  {author} {\bibfnamefont {B.~M.}\ \bibnamefont {Nyman}},\ }\href@noop {}
  {\bibfield  {journal} {\bibinfo  {journal} {Electronic Lett.}\ }\textbf
  {\bibinfo {volume} {27}},\ \bibinfo {pages} {2055} (\bibinfo {year}
  {1991}{\natexlab{b}})}\BibitemShut {NoStop}%
\bibitem [{\citenamefont {Haus}\ and\ \citenamefont {Wong}(1996)}]{Haus_1996}%
  \BibitemOpen
  \bibfield  {author} {\bibinfo {author} {\bibfnamefont {H.~A.}\ \bibnamefont
  {Haus}}\ and\ \bibinfo {author} {\bibfnamefont {W.~S.}\ \bibnamefont
  {Wong}},\ }\href@noop {} {\bibfield  {journal} {\bibinfo  {journal} {Rev.
  Mod. Phys.}\ }\textbf {\bibinfo {volume} {68}},\ \bibinfo {pages} {423}
  (\bibinfo {year} {1996})}\BibitemShut {NoStop}%
\bibitem [{\citenamefont {Husakou}\ and\ \citenamefont
  {Herrmann}(2001)}]{Husakou_2001}%
  \BibitemOpen
  \bibfield  {author} {\bibinfo {author} {\bibfnamefont {A.~V.}\ \bibnamefont
  {Husakou}}\ and\ \bibinfo {author} {\bibfnamefont {J.}~\bibnamefont
  {Herrmann}},\ }\href@noop {} {\bibfield  {journal} {\bibinfo  {journal}
  {Phys. Rev. Lett.}\ }\textbf {\bibinfo {volume} {87}},\ \bibinfo {pages}
  {203901} (\bibinfo {year} {2001})}\BibitemShut {NoStop}%
\bibitem [{\citenamefont {Dudley}\ \emph {et~al.}(2006)\citenamefont {Dudley},
  \citenamefont {Genty},\ and\ \citenamefont {Coen}}]{Dudley_2006}%
  \BibitemOpen
  \bibfield  {author} {\bibinfo {author} {\bibfnamefont {J.~M.}\ \bibnamefont
  {Dudley}}, \bibinfo {author} {\bibfnamefont {G.}~\bibnamefont {Genty}}, \
  and\ \bibinfo {author} {\bibfnamefont {S.}~\bibnamefont {Coen}},\ }\href@noop
  {} {\bibfield  {journal} {\bibinfo  {journal} {Rev. Mod. Phys.}\ }\textbf
  {\bibinfo {volume} {78}},\ \bibinfo {pages} {1135} (\bibinfo {year}
  {2006})}\BibitemShut {NoStop}%
\bibitem [{\citenamefont {Xu}\ and\ \citenamefont {Wise}(2013)}]{Xu_2013}%
  \BibitemOpen
  \bibfield  {author} {\bibinfo {author} {\bibfnamefont {C.}~\bibnamefont
  {Xu}}\ and\ \bibinfo {author} {\bibfnamefont {F.~W.}\ \bibnamefont {Wise}},\
  }\href@noop {} {\bibfield  {journal} {\bibinfo  {journal} {Nat. Photonics}\
  }\textbf {\bibinfo {volume} {7}},\ \bibinfo {pages} {875} (\bibinfo {year}
  {2013})}\BibitemShut {NoStop}%
\bibitem [{\citenamefont {Cundiff}\ and\ \citenamefont
  {Ye}(2003)}]{Cundiff_2003}%
  \BibitemOpen
  \bibfield  {author} {\bibinfo {author} {\bibfnamefont {S.~T.}\ \bibnamefont
  {Cundiff}}\ and\ \bibinfo {author} {\bibfnamefont {J.}~\bibnamefont {Ye}},\
  }\href@noop {} {\bibfield  {journal} {\bibinfo  {journal} {Rev. Mod. Phys.}\
  }\textbf {\bibinfo {volume} {75}},\ \bibinfo {pages} {325} (\bibinfo {year}
  {2003})}\BibitemShut {NoStop}%
\bibitem [{\citenamefont {Zhou}\ \emph {et~al.}(1994)\citenamefont {Zhou},
  \citenamefont {Taft}, \citenamefont {Huang}, \citenamefont {Murnane},
  \citenamefont {Kapteyn},\ and\ \citenamefont {Christov}}]{Zhou_1994}%
  \BibitemOpen
  \bibfield  {author} {\bibinfo {author} {\bibfnamefont {J.}~\bibnamefont
  {Zhou}}, \bibinfo {author} {\bibfnamefont {G.}~\bibnamefont {Taft}}, \bibinfo
  {author} {\bibfnamefont {C.~P.}\ \bibnamefont {Huang}}, \bibinfo {author}
  {\bibfnamefont {M.~M.}\ \bibnamefont {Murnane}}, \bibinfo {author}
  {\bibfnamefont {H.~C.}\ \bibnamefont {Kapteyn}}, \ and\ \bibinfo {author}
  {\bibfnamefont {I.~P.}\ \bibnamefont {Christov}},\ }\href@noop {} {\bibfield
  {journal} {\bibinfo  {journal} {Opt. Lett.}\ }\textbf {\bibinfo {volume}
  {19}},\ \bibinfo {pages} {1149} (\bibinfo {year} {1994})}\BibitemShut
  {NoStop}%
\bibitem [{\citenamefont {Jung}\ \emph {et~al.}(1997)\citenamefont {Jung},
  \citenamefont {K{\"{a}}rtner}, \citenamefont {Matuschek}, \citenamefont
  {Sutter}, \citenamefont {Morier-Genoud}, \citenamefont {Zhang}, \citenamefont
  {Keller}, \citenamefont {Scheuer}, \citenamefont {Tilsch},\ and\
  \citenamefont {Tschudi}}]{Jung_1997}%
  \BibitemOpen
  \bibfield  {author} {\bibinfo {author} {\bibfnamefont {I.~D.}\ \bibnamefont
  {Jung}}, \bibinfo {author} {\bibfnamefont {F.~X.}\ \bibnamefont
  {K{\"{a}}rtner}}, \bibinfo {author} {\bibfnamefont {N.}~\bibnamefont
  {Matuschek}}, \bibinfo {author} {\bibfnamefont {D.~H.}\ \bibnamefont
  {Sutter}}, \bibinfo {author} {\bibfnamefont {F.}~\bibnamefont
  {Morier-Genoud}}, \bibinfo {author} {\bibfnamefont {G.}~\bibnamefont
  {Zhang}}, \bibinfo {author} {\bibfnamefont {U.}~\bibnamefont {Keller}},
  \bibinfo {author} {\bibfnamefont {V.}~\bibnamefont {Scheuer}}, \bibinfo
  {author} {\bibfnamefont {M.}~\bibnamefont {Tilsch}}, \ and\ \bibinfo {author}
  {\bibfnamefont {T.}~\bibnamefont {Tschudi}},\ }\href@noop {} {\bibfield
  {journal} {\bibinfo  {journal} {Opt. Lett.}\ }\textbf {\bibinfo {volume}
  {22}},\ \bibinfo {pages} {1009} (\bibinfo {year} {1997})}\BibitemShut
  {NoStop}%
\bibitem [{\citenamefont {Mollenauer}\ and\ \citenamefont
  {Stolen}(1984)}]{Mollenauer_1984}%
  \BibitemOpen
  \bibfield  {author} {\bibinfo {author} {\bibfnamefont {L.~F.}\ \bibnamefont
  {Mollenauer}}\ and\ \bibinfo {author} {\bibfnamefont {R.~H.}\ \bibnamefont
  {Stolen}},\ }\href@noop {} {\bibfield  {journal} {\bibinfo  {journal} {Opt.
  Lett.}\ }\textbf {\bibinfo {volume} {9}},\ \bibinfo {pages} {13} (\bibinfo
  {year} {1984})}\BibitemShut {NoStop}%
\bibitem [{\citenamefont {Kafka}\ \emph {et~al.}(1989)\citenamefont {Kafka},
  \citenamefont {Baer},\ and\ \citenamefont {Hall}}]{Kafka_1989}%
  \BibitemOpen
  \bibfield  {author} {\bibinfo {author} {\bibfnamefont {J.~D.}\ \bibnamefont
  {Kafka}}, \bibinfo {author} {\bibfnamefont {T.}~\bibnamefont {Baer}}, \ and\
  \bibinfo {author} {\bibfnamefont {D.~W.}\ \bibnamefont {Hall}},\ }\href@noop
  {} {\bibfield  {journal} {\bibinfo  {journal} {Opt. Lett.}\ }\textbf
  {\bibinfo {volume} {14}},\ \bibinfo {pages} {1269} (\bibinfo {year}
  {1989})}\BibitemShut {NoStop}%
\bibitem [{\citenamefont {Matsas}\ \emph {et~al.}(1992)\citenamefont {Matsas},
  \citenamefont {Newson}, ,\ and\ \citenamefont {Zervas}}]{Matsas_1992}%
  \BibitemOpen
  \bibfield  {author} {\bibinfo {author} {\bibfnamefont {V.~J.}\ \bibnamefont
  {Matsas}}, \bibinfo {author} {\bibfnamefont {T.~P.}\ \bibnamefont {Newson}},
  , \ and\ \bibinfo {author} {\bibfnamefont {M.~N.}\ \bibnamefont {Zervas}},\
  }\href@noop {} {\bibfield  {journal} {\bibinfo  {journal} {Opt. Comm.}\
  }\textbf {\bibinfo {volume} {92}},\ \bibinfo {pages} {61} (\bibinfo {year}
  {1992})}\BibitemShut {NoStop}%
\bibitem [{\citenamefont {Zakharov}\ and\ \citenamefont
  {Shabat}(1972)}]{Zakharov_1972}%
  \BibitemOpen
  \bibfield  {author} {\bibinfo {author} {\bibfnamefont {V.~E.}\ \bibnamefont
  {Zakharov}}\ and\ \bibinfo {author} {\bibfnamefont {A.~B.}\ \bibnamefont
  {Shabat}},\ }\href@noop {} {\bibfield  {journal} {\bibinfo  {journal} {Sov.
  Phys. JETP}\ }\textbf {\bibinfo {volume} {34}},\ \bibinfo {pages} {62}
  (\bibinfo {year} {1972})}\BibitemShut {NoStop}%
\bibitem [{\citenamefont {Hasegawa}\ and\ \citenamefont
  {Tappert}(1973)}]{Hasegawa_1973}%
  \BibitemOpen
  \bibfield  {author} {\bibinfo {author} {\bibfnamefont {A.}~\bibnamefont
  {Hasegawa}}\ and\ \bibinfo {author} {\bibfnamefont {F.}~\bibnamefont
  {Tappert}},\ }\href@noop {} {\bibfield  {journal} {\bibinfo  {journal} {Appl.
  Phys. Lett.}\ }\textbf {\bibinfo {volume} {23}},\ \bibinfo {pages} {142}
  (\bibinfo {year} {1973})}\BibitemShut {NoStop}%
\bibitem [{\citenamefont {Kelly}(1992)}]{Kelly_1992}%
  \BibitemOpen
  \bibfield  {author} {\bibinfo {author} {\bibfnamefont {S.~M.~J.}\
  \bibnamefont {Kelly}},\ }\href@noop {} {\bibfield  {journal} {\bibinfo
  {journal} {Electronic Lett.}\ }\textbf {\bibinfo {volume} {28}},\ \bibinfo
  {pages} {806} (\bibinfo {year} {1992})}\BibitemShut {NoStop}%
\bibitem [{\citenamefont {Blanco-Redondo}\ \emph {et~al.}(2016)\citenamefont
  {Blanco-Redondo}, \citenamefont {de~Sterke}, \citenamefont {Sipe},
  \citenamefont {Krauss}, \citenamefont {Eggleton}, ,\ and\ \citenamefont
  {Husko}}]{Blanco_Redondo_2016}%
  \BibitemOpen
  \bibfield  {author} {\bibinfo {author} {\bibfnamefont {A.}~\bibnamefont
  {Blanco-Redondo}}, \bibinfo {author} {\bibfnamefont {C.~M.}\ \bibnamefont
  {de~Sterke}}, \bibinfo {author} {\bibfnamefont {J.~E.}\ \bibnamefont {Sipe}},
  \bibinfo {author} {\bibfnamefont {T.~F.}\ \bibnamefont {Krauss}}, \bibinfo
  {author} {\bibfnamefont {B.~J.}\ \bibnamefont {Eggleton}}, , \ and\ \bibinfo
  {author} {\bibfnamefont {C.}~\bibnamefont {Husko}},\ }\href@noop {}
  {\bibfield  {journal} {\bibinfo  {journal} {Nature Comm.}\ }\textbf {\bibinfo
  {volume} {7}},\ \bibinfo {pages} {10427} (\bibinfo {year}
  {2016})}\BibitemShut {NoStop}%
\bibitem [{\citenamefont {Lo}\ \emph {et~al.}(2018)\citenamefont {Lo},
  \citenamefont {Stefani}, \citenamefont {de~Sterke},\ and\ \citenamefont
  {Blanco-Redondo}}]{Lo_2018}%
  \BibitemOpen
  \bibfield  {author} {\bibinfo {author} {\bibfnamefont {C.~W.}\ \bibnamefont
  {Lo}}, \bibinfo {author} {\bibfnamefont {A.}~\bibnamefont {Stefani}},
  \bibinfo {author} {\bibfnamefont {C.~M.}\ \bibnamefont {de~Sterke}}, \ and\
  \bibinfo {author} {\bibfnamefont {A.}~\bibnamefont {Blanco-Redondo}},\
  }\href@noop {} {\bibfield  {journal} {\bibinfo  {journal} {Opt. Express}\
  }\textbf {\bibinfo {volume} {26}},\ \bibinfo {pages} {7786} (\bibinfo {year}
  {2018})}\BibitemShut {NoStop}%
\bibitem [{\citenamefont {Tam}\ \emph {et~al.}(2019)\citenamefont {Tam},
  \citenamefont {Alexander}, \citenamefont {Blanco-Redondo},\ and\
  \citenamefont {de~Sterke}}]{Tam_2019}%
  \BibitemOpen
  \bibfield  {author} {\bibinfo {author} {\bibfnamefont {K.~K.~K.}\
  \bibnamefont {Tam}}, \bibinfo {author} {\bibfnamefont {T.}~\bibnamefont
  {Alexander}}, \bibinfo {author} {\bibfnamefont {A.}~\bibnamefont
  {Blanco-Redondo}}, \ and\ \bibinfo {author} {\bibfnamefont {C.~M.}\
  \bibnamefont {de~Sterke}},\ }\href@noop {} {\bibfield  {journal} {\bibinfo
  {journal} {Opt. Lett.}\ }\textbf {\bibinfo {volume} {44}},\ \bibinfo {pages}
  {3306} (\bibinfo {year} {2019})}\BibitemShut {NoStop}%
\bibitem [{\citenamefont {Taheri}\ and\ \citenamefont
  {Matsko}(2019)}]{TaheriMatsko_2019}%
  \BibitemOpen
  \bibfield  {author} {\bibinfo {author} {\bibfnamefont {H.}~\bibnamefont
  {Taheri}}\ and\ \bibinfo {author} {\bibfnamefont {A.~B.}\ \bibnamefont
  {Matsko}},\ }\href@noop {} {\bibfield  {journal} {\bibinfo  {journal} {Opt.
  Lett.}\ }\textbf {\bibinfo {volume} {44}},\ \bibinfo {pages} {3086} (\bibinfo
  {year} {2019})}\BibitemShut {NoStop}%
\bibitem [{\citenamefont {Knox}\ \emph {et~al.}(1994)\citenamefont {Knox},
  \citenamefont {Forysiak},\ and\ \citenamefont {Doran}}]{Knox_1994}%
  \BibitemOpen
  \bibfield  {author} {\bibinfo {author} {\bibfnamefont {F.}~\bibnamefont
  {Knox}}, \bibinfo {author} {\bibfnamefont {W.}~\bibnamefont {Forysiak}}, \
  and\ \bibinfo {author} {\bibfnamefont {N.}~\bibnamefont {Doran}},\
  }\href@noop {} {\bibfield  {journal} {\bibinfo  {journal} {IEEE J. Lightwave
  Technol.}\ }\textbf {\bibinfo {volume} {13}},\ \bibinfo {pages} {1955–}
  (\bibinfo {year} {1994})}\BibitemShut {NoStop}%
\bibitem [{\citenamefont {Turitsyn}\ \emph {et~al.}(2012)\citenamefont
  {Turitsyn}, \citenamefont {Bale},\ and\ \citenamefont
  {Fedoruk}}]{Turitsyn_2012}%
  \BibitemOpen
  \bibfield  {author} {\bibinfo {author} {\bibfnamefont {S.~K.}\ \bibnamefont
  {Turitsyn}}, \bibinfo {author} {\bibfnamefont {B.~G.}\ \bibnamefont {Bale}},
  \ and\ \bibinfo {author} {\bibfnamefont {M.~P.}\ \bibnamefont {Fedoruk}},\
  }\href@noop {} {\bibfield  {journal} {\bibinfo  {journal} {Physics Reports}\
  }\textbf {\bibinfo {volume} {521}},\ \bibinfo {pages} {135} (\bibinfo {year}
  {2012})}\BibitemShut {NoStop}%
\bibitem [{\citenamefont {Agrawal}(1995)}]{Agrawal_NFO}%
  \BibitemOpen
  \bibfield  {author} {\bibinfo {author} {\bibfnamefont {G.~P.}\ \bibnamefont
  {Agrawal}},\ }\enquote {\bibinfo {title} {Nonlinear fibre optics},}\ \
  (\bibinfo  {publisher} {Academic Press},\ \bibinfo {year} {1995})\ \bibinfo
  {edition} {2nd}\ ed.\BibitemShut {Stop}%
\bibitem [{\citenamefont {Schr{\"{o}}der}\ \emph {et~al.}(2010)\citenamefont
  {Schr{\"{o}}der}, \citenamefont {Coen}, \citenamefont {Sylvestre},\ and\
  \citenamefont {Eggleton}}]{Schroder_2010}%
  \BibitemOpen
  \bibfield  {author} {\bibinfo {author} {\bibfnamefont {J.}~\bibnamefont
  {Schr{\"{o}}der}}, \bibinfo {author} {\bibfnamefont {S.}~\bibnamefont
  {Coen}}, \bibinfo {author} {\bibfnamefont {T.}~\bibnamefont {Sylvestre}}, \
  and\ \bibinfo {author} {\bibfnamefont {B.~J.}\ \bibnamefont {Eggleton}},\
  }\href@noop {} {\bibfield  {journal} {\bibinfo  {journal} {Opt. Express}\
  }\textbf {\bibinfo {volume} {18}},\ \bibinfo {pages} {22715} (\bibinfo {year}
  {2010})}\BibitemShut {NoStop}%
\bibitem [{\citenamefont {Peng}\ and\ \citenamefont
  {Boscolo}(2016)}]{Peng_2016}%
  \BibitemOpen
  \bibfield  {author} {\bibinfo {author} {\bibfnamefont {J.}~\bibnamefont
  {Peng}}\ and\ \bibinfo {author} {\bibfnamefont {S.}~\bibnamefont {Boscolo}},\
  }\href@noop {} {\bibfield  {journal} {\bibinfo  {journal} {Sci. Rep.}\
  }\textbf {\bibinfo {volume} {6}},\ \bibinfo {pages} {25995} (\bibinfo {year}
  {2016})}\BibitemShut {NoStop}%
\bibitem [{\citenamefont {Dorrer}\ and\ \citenamefont
  {Kang}(2002)}]{Dorrer_2002}%
  \BibitemOpen
  \bibfield  {author} {\bibinfo {author} {\bibfnamefont {C.}~\bibnamefont
  {Dorrer}}\ and\ \bibinfo {author} {\bibfnamefont {I.}~\bibnamefont {Kang}},\
  }\href@noop {} {\bibfield  {journal} {\bibinfo  {journal} {Opt. Lett.}\
  }\textbf {\bibinfo {volume} {27}},\ \bibinfo {pages} {1315} (\bibinfo {year}
  {2002})}\BibitemShut {NoStop}%
\bibitem [{\citenamefont {Hammani}\ \emph {et~al.}(2011)\citenamefont
  {Hammani}, \citenamefont {Kibler}, \citenamefont {Finot}, \citenamefont
  {Morin}, \citenamefont {Fatome}, \citenamefont {Dudley}, ,\ and\
  \citenamefont {Millot}}]{Hammani_2011}%
  \BibitemOpen
  \bibfield  {author} {\bibinfo {author} {\bibfnamefont {K.}~\bibnamefont
  {Hammani}}, \bibinfo {author} {\bibfnamefont {B.}~\bibnamefont {Kibler}},
  \bibinfo {author} {\bibfnamefont {C.}~\bibnamefont {Finot}}, \bibinfo
  {author} {\bibfnamefont {P.}~\bibnamefont {Morin}}, \bibinfo {author}
  {\bibfnamefont {J.}~\bibnamefont {Fatome}}, \bibinfo {author} {\bibfnamefont
  {J.~M.}\ \bibnamefont {Dudley}}, , \ and\ \bibinfo {author} {\bibfnamefont
  {G.}~\bibnamefont {Millot}},\ }\href@noop {} {\bibfield  {journal} {\bibinfo
  {journal} {Opt. Lett.}\ }\textbf {\bibinfo {volume} {36}},\ \bibinfo {pages}
  {112} (\bibinfo {year} {2011})}\BibitemShut {NoStop}%
\bibitem [{\citenamefont {Ito}\ \emph {et~al.}(2016)\citenamefont {Ito},
  \citenamefont {Slezak}, \citenamefont {Yoshita}, \citenamefont {Akiyama},\
  and\ \citenamefont {Kobayashi}}]{Ito_2016}%
  \BibitemOpen
  \bibfield  {author} {\bibinfo {author} {\bibfnamefont {T.}~\bibnamefont
  {Ito}}, \bibinfo {author} {\bibfnamefont {O.}~\bibnamefont {Slezak}},
  \bibinfo {author} {\bibfnamefont {M.}~\bibnamefont {Yoshita}}, \bibinfo
  {author} {\bibfnamefont {H.}~\bibnamefont {Akiyama}}, \ and\ \bibinfo
  {author} {\bibfnamefont {Y.}~\bibnamefont {Kobayashi}},\ }\href@noop {}
  {\bibfield  {journal} {\bibinfo  {journal} {Photon. Research}\ }\textbf
  {\bibinfo {volume} {4}},\ \bibinfo {pages} {13} (\bibinfo {year}
  {2016})}\BibitemShut {NoStop}%
\bibitem [{\citenamefont {Oktem}\ \emph {et~al.}(2010)\citenamefont {Oktem},
  \citenamefont {{\"{U}}lg{\"{u}}d{\"{u}}r},\ and\ \citenamefont
  {Ilday}}]{Oktem_2010}%
  \BibitemOpen
  \bibfield  {author} {\bibinfo {author} {\bibfnamefont {B.}~\bibnamefont
  {Oktem}}, \bibinfo {author} {\bibfnamefont {C.}~\bibnamefont
  {{\"{U}}lg{\"{u}}d{\"{u}}r}}, \ and\ \bibinfo {author} {\bibfnamefont
  {F.~{\"{O}}.}\ \bibnamefont {Ilday}},\ }\href@noop {} {\bibfield  {journal}
  {\bibinfo  {journal} {Nat. Photonics}\ }\textbf {\bibinfo {volume} {4}},\
  \bibinfo {pages} {307} (\bibinfo {year} {2010})}\BibitemShut {NoStop}%
\bibitem [{\citenamefont {Woodward}(2018)}]{Woodward_2018}%
  \BibitemOpen
  \bibfield  {author} {\bibinfo {author} {\bibfnamefont {R.~I.}\ \bibnamefont
  {Woodward}},\ }\href@noop {} {\bibfield  {journal} {\bibinfo  {journal} {J.
  Opt.}\ }\textbf {\bibinfo {volume} {20}},\ \bibinfo {pages} {033002}
  (\bibinfo {year} {2018})}\BibitemShut {NoStop}%
\bibitem [{\citenamefont {Dennis}\ and\ \citenamefont
  {Duling}(1994)}]{Dennis_1994}%
  \BibitemOpen
  \bibfield  {author} {\bibinfo {author} {\bibfnamefont {M.~L.}\ \bibnamefont
  {Dennis}}\ and\ \bibinfo {author} {\bibfnamefont {I.~N.}\ \bibnamefont
  {Duling}},\ }\href@noop {} {\bibfield  {journal} {\bibinfo  {journal} {IEEE
  J. Quant. Elect.}\ }\textbf {\bibinfo {volume} {30}},\ \bibinfo {pages}
  {1469} (\bibinfo {year} {1994})}\BibitemShut {NoStop}%
\bibitem [{\citenamefont {Tamura}\ \emph {et~al.}(1993)\citenamefont {Tamura},
  \citenamefont {Ippen}, \citenamefont {Haus},\ and\ \citenamefont
  {Nelson}}]{Tamura_1993}%
  \BibitemOpen
  \bibfield  {author} {\bibinfo {author} {\bibfnamefont {K.}~\bibnamefont
  {Tamura}}, \bibinfo {author} {\bibfnamefont {E.~P.}\ \bibnamefont {Ippen}},
  \bibinfo {author} {\bibfnamefont {H.~A.}\ \bibnamefont {Haus}}, \ and\
  \bibinfo {author} {\bibfnamefont {L.~E.}\ \bibnamefont {Nelson}},\
  }\href@noop {} {\bibfield  {journal} {\bibinfo  {journal} {Opt. Lett.}\
  }\textbf {\bibinfo {volume} {18}},\ \bibinfo {pages} {1080} (\bibinfo {year}
  {1993})}\BibitemShut {NoStop}%
\bibitem [{\citenamefont {Chen}\ \emph {et~al.}(1999)\citenamefont {Chen},
  \citenamefont {K{\"{a}}rtner}, \citenamefont {Morgner}, \citenamefont {Cho},
  \citenamefont {Haus}, \citenamefont {Ippen},\ and\ \citenamefont
  {Fujimoto}}]{Chen_1999}%
  \BibitemOpen
  \bibfield  {author} {\bibinfo {author} {\bibfnamefont {Y.}~\bibnamefont
  {Chen}}, \bibinfo {author} {\bibfnamefont {F.~X.}\ \bibnamefont
  {K{\"{a}}rtner}}, \bibinfo {author} {\bibfnamefont {U.}~\bibnamefont
  {Morgner}}, \bibinfo {author} {\bibfnamefont {S.~H.}\ \bibnamefont {Cho}},
  \bibinfo {author} {\bibfnamefont {H.~A.}\ \bibnamefont {Haus}}, \bibinfo
  {author} {\bibfnamefont {E.~P.}\ \bibnamefont {Ippen}}, \ and\ \bibinfo
  {author} {\bibfnamefont {J.~G.}\ \bibnamefont {Fujimoto}},\ }\href@noop {}
  {\bibfield  {journal} {\bibinfo  {journal} {J. Opt. Soc. Am. B}\ }\textbf
  {\bibinfo {volume} {16}},\ \bibinfo {pages} {1999} (\bibinfo {year}
  {1999})}\BibitemShut {NoStop}%
\bibitem [{\citenamefont {Blanco-Redondo}\ \emph {et~al.}(2017)\citenamefont
  {Blanco-Redondo}, \citenamefont {de~Sterke}, \citenamefont {Husko},\ and\
  \citenamefont {Eggleton}}]{Blanco_Redondo_2017}%
  \BibitemOpen
  \bibfield  {author} {\bibinfo {author} {\bibfnamefont {A.}~\bibnamefont
  {Blanco-Redondo}}, \bibinfo {author} {\bibfnamefont {C.~M.}\ \bibnamefont
  {de~Sterke}}, \bibinfo {author} {\bibfnamefont {C.}~\bibnamefont {Husko}}, \
  and\ \bibinfo {author} {\bibfnamefont {B.~J.}\ \bibnamefont {Eggleton}},\
  }in\ \href@noop {} {\emph {\bibinfo {booktitle} {European Quantum Electronics
  Conference, paper EE\_3\_2}}}\ (\bibinfo  {publisher} {Optical Society of
  America},\ \bibinfo {year} {2017})\BibitemShut {NoStop}%
\end{thebibliography}%

\end{document}